\documentstyle[prd,aps,psfig,epsfig]{revtex} 
\def\lsim{\raise0.3ex\hbox{$\;<$\kern-0.75em\raise-1.1ex\hbox{$\sim\;$}}}
\def\gsim{\raise0.3ex\hbox{$\;>$\kern-0.75em\raise-1.1ex\hbox{$\sim\;$}}}

\def\vev#1{\langle #1\rangle}
\def\cl{$^{37}$Cl\ }
\def\ga{$^{71}$Ga\  }
\def\be7{$^{7}$Be}
\def\b8{$^{8}$B}
\draft

\begin{document}

\preprint{}
\draft

\title{
Current Status of the Resonant Spin-Flavor Precession Solution 
to the Solar Neutrino Problem}

\author{M. M. Guzzo\thanks{E-mail: guzzo@ifi.unicamp.br}
and H. Nunokawa\thanks{E-mail: nunokawa@ifi.unicamp.br}}
\address{
Instituto de F\'{\i}sica Gleb Wataghin\\
Universidade Estadual de Campinas - UNICAMP\\
13083-970 Campinas SP Brasil
}
\date{October, 1998, Revised March, 1999}

\maketitle
\begin{abstract}
We discuss the current status of the resonant spin-flavor 
precession (RSFP) solution to the solar neutrino problem.
We perform a fit to all the latest solar neutrino data
for various assumed magnetic field profiles in the Sun. 
We show that the RSFP can account for all the solar neutrino 
experiments, giving as good fit as other alternative 
solutions such as MSW or vacuum oscillation (Just So), 
and therefore can be a viable solution to the solar neutrino problem 
\end{abstract}

\vskip 0.5cm

\section{Introduction}
The solar neutrino anomaly has been more and more strongly
established both in 
its experimental~\cite{homestake,kamiokande,sage,gallex,superkam} 
as well as in its theoretical~\cite{JBahcall,BP95,BP98,reviewSSM} aspect. 
In fact, both have presented a  very dynamical evolution. 
{}From one side, theoretical predictions, i.e., the standard 
solar models (SSM),  have been refined  by including several 
mechanisms  such as helium diffusion~\cite{BP95,BP98} and 
an updating analysis of the $S_{17}$ astrophysical 
factor~\cite{BP98,INT}. 
One can see in ref. \cite{BKS} that theoretical predictions obtained by 
different SSMs, which are developed independently,    
are in good agreement with each other. 
Moreover, it has been shown that the SSM is in excellent 
agreement with  the helioseismology \cite{BP98}. 

On the other hand, experimental data have become more accurate
due to the calibration of the experiments, more
statistics and the existence of the new generation SuperKamiokande experiment
and its solar neutrino spectral observations~\cite{superkam}.  Consequently,
final numbers related to the solar neutrino deficit have also evolved
significantly. The most updated experimental data as well as 
theoretical predictions are shown in Table I. 
One can show that these observed solar neutrino data are in strong 
disagreement with the ones predicted by the SSM \cite{BKS,minakata98}.  
Moreover, this conclusion does not depend on any details of the SSM.  

Solutions to the solar neutrino problem rely on different phenomena
~\cite{30years} which deplete the number of observable neutrinos 
at the Earth: neutrino oscillations in vacuum~\cite{vacuum},
resonantly enhanced matter oscillations~\cite{msw}, resonant spin-flavor
precession phenomenon (RSFP)~\cite{LimMarciano,Akhmedov} 
and flavor mixing by new interactions~\cite{fcnc1,fcnc2}. 
The capability of each one of these processes to make compatible solar 
neutrino predictions and observations and therefore to find a solution 
to the solar neutrino anomaly have been updated from time to 
time~\cite{update}. 
In this paper we investigate the current status of the RSFP 
scenario \cite{review}. 
We believe that it is worthwhile to reanalyse this mechanism 
in the light of new solar neutrino data as well as the new SSM.
We also discuss briefly how the solar neutrino spectral observations 
in SuperKamiokande are affected by the RSFP mechanism.
\vglue 0.5cm
\begin{table}[th]
\label{tab:data}
\caption{Observed solar neutrino event rates used in this analysis and 
corresponding predictions from the reference standard solar model
\protect\cite{BP98}. The quoted errors are at $1\sigma$.}
\begin{tabular}{ccccc}

Experiment 	& Data~$\pm$(stat.)~$\pm$(syst.)&Ref.
& Theory \protect\cite{BP98}& Units \\
\tableline
Homestake	& $ 2.56 \pm 0.16 \pm 0.15$        & \protect\cite{homestake} &
	$7.7^{+1.2}_{-1.0}$    & SNU					\\
SAGE		&$69.9^{+8.0}_{-7.7}{}^{+3.9}_{-4.1}$& \protect\cite{sage}&
	$129^{+8}_{-6}$        & SNU					\\
GALLEX		& $76.4 \pm  6.3^{+4.5}_{-4.9}$     & \protect\cite{gallex}&
	 $129^{+8}_{-6}$       & SNU					\\
SuperKamiokande	&$2.44 \pm +0.05{}^{+0.09}_{-0.06}$& \protect\cite{superkam}&
 	$5.15^{+0.98}_{-0.72}$  &  $10^6$ cm$^{-2}$s$^{-1}$	
\end{tabular}
\vglue -0.5cm
\end{table}
RSFP mechanism is very sensitive to the magnetic profile in the Sun and, 
in fact, several possible scenarios for the magnetic strength have been 
proposed by different authors \cite{alp1,krastev,limnunokawa,pulido,chauhan}.
We consider several possibilities which include the main qualitative
aspects of the magnetic profile in the Sun previously invoked as a solution 
to the solar neutrino anomaly. 
Using the minimum $\chi^2$ method to compare theoretical predictions of 
the RSFP phenomenon and the experimental observations we conclude that 
very good fits can be obtained for the average solar neutrino suppression, 
if intense magnetic fields in the solar convective zone are considered. 

\section{RSFP mechanism}

Assuming a nonvanishing transition magnetic moment of neutrinos, active
solar neutrinos interacting with the magnetic field in the Sun can be
spin-flavor converted into sterile nonelectron neutrinos~\cite{cisneros,vvo} 
(if we are dealing with Dirac particles)
or into active nonelectron antineutrinos~\cite{schechtervalle} (if the involved
particles are Majorana).  In both cases the resulting particles
interact with solar neutrino detectors significantly less than the
original active electron neutrinos in such a way that this phenomenon
can induce a depletion in the detectable solar neutrino flux.

Spin-flavor precession of neutrinos can be resonantly enhanced in
matter~\cite{LimMarciano,Akhmedov}, in close analogy with the MSW
effect~\cite{msw}. In this case the precession strongly depends 
on the neutrino energy and provokes different suppressions for 
each portion of the solar neutrino energy spectrum. 
Therefore RSFP provides a
satisfactory description~\cite{review,alp1,krastev,limnunokawa,pulido,chauhan} 
of the actual experimental
panorama~\cite{homestake,kamiokande,sage,gallex,superkam}: all experiments
detect less than the theoretically predicted solar neutrino
fluxes~\cite{BP98} and  different suppressions are observed in
each experiment, suggesting that the mechanism to conciliate
theoretical predictions and observations has to differentiate the
different parts of the solar neutrino spectrum.

For simplicity, we consider two generation of neutrinos, electron neutrino 
and, for e.g., muon neutrino (which could be replaced by tau neutrino 
in our discussion). 
Furthermore, we assume that the vacuum mixing angle is zero or small enough 
to be neglected. (See ref. \cite{rsfpmsw} for the case where RSFP and flavor 
mixing simultaneously exists.) 
The time evolution of neutrinos interacting with a magnetic field $B$ 
through a nonvanishing neutrino magnetic moment $\mu_\nu$ in matter 
is governed by a Schr\"odinger-like equation \cite{LimMarciano,Akhmedov};

\begin{equation}
i \frac{d}{dt}
\left(
\begin{array}{l}
\nu_{e_L} \\
\overline{\nu}_{\mu_R}
\end{array}
\right)
=
\left(
\begin{array}{cl}
a_{\nu_e} \,  \,   &   \mu_\nu B    \\
\mu_\nu B     \,  \,   &   \frac{\Delta m^2}{2 E} + a_{\nu_{\mu}}
\end{array}
\right)
\left(
\begin{array}{l}
\nu_{e_L}   \\
\overline{\nu}_{\mu_R}
\end{array}
\right)  ,
\label{evolution}
\end{equation}
where $\nu_e$  and $\overline{\nu}_{\mu_R}$ are active electron
neutrinos and muon antineutrinos, respectively, 
$\Delta m^2= m^2_{\nu_\mu} -m^2_{\nu_e}$ is their squared mass difference and 
$E$ is the neutrino energy, $a_{\nu_e}= G_F(2N_e-N_n)/\sqrt{2}$ and
$a_{\nu_\mu}=G_FN_n/\sqrt{2}$, with $N_e$ and $N_n$ being electron and
neutron number densities, respectively. 
In eq. (\ref{evolution}) we are assuming that 
neutrinos are Majorana particles. 
For the Dirac case, the spin-flavor  precession involves
$\nu_e \leftrightarrow \nu_s$, where $\nu_s$ is a sterile neutrino
and $a_{\nu_s}=0$.
%
%
\begin{figure}[ht]
\vglue -1.2cm
\centerline{
\psfig{file=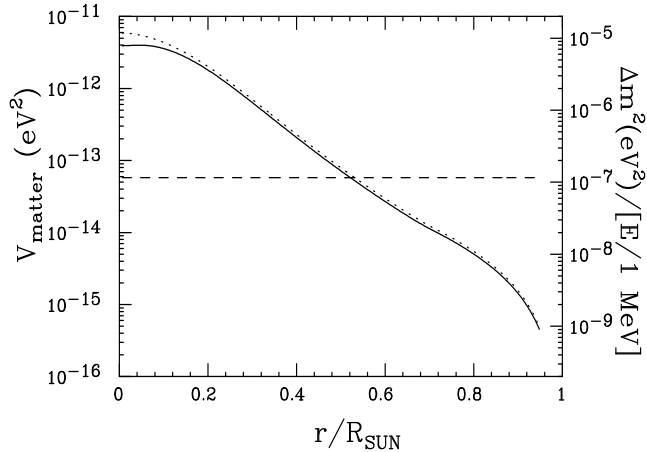,height=9.4cm,width=7.8cm,angle=90}}
\vglue -0.5cm
\caption{
Matter potentials as a function of radial 
distance from the solar center are plotted. 
The solid and dotted curves correspond to 
the Majorana and Dirac case, respectively. 
The dashed line correspond to 
$\mu_\nu B = 10^{-11}\mu_B\times 100$ kG. 
\label{fig:mattpotential}}
\end{figure}

\section{Analysis}

In order to obtain the survival probability
we first numerically integrate the evolution equations~(\ref{evolution})
varying matter density in the Sun \cite{JBahcall} 
for some assumed profiles of the magnetic field which 
will be described below. 
Next, using the solar neutrino flux in ref. \cite{BP98}, 
we compute the expected solar neutrino event rate in each experiment, 
taking into account the relevant absorption
cross sections~\cite{JBahcall}  for \ga and \cl experiments 
as well as the scattering cross sections for 
$\nu_e$-$e^-$ and $\bar{\nu}_\mu$-$e^-$ reactions including 
also the efficiency function for the SuperKamiokande experiment 
in the same way as in ref. \cite{BKL}.
We note that in this analysis we always adopt the 
solar model in ref. \cite{BP98} as a 
reference SSM. 
%
\begin{figure}[ht]
\vglue -0.5cm
\centerline{
\epsfig{file=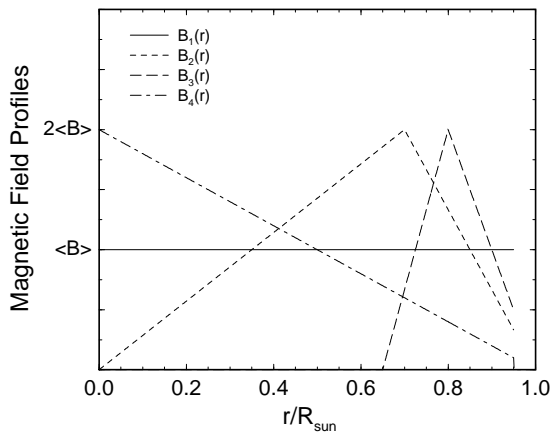,width=7.7cm} }
\noindent
\caption{
Various magnetic field profiles used in this work. 
For each field 
$\vev{B}$ is defined as the 
average of the field over the region where 
$B(r)$ is not zero. 
\label{fig:profile}}
\vglue -0.5cm
\end{figure}

\subsection{Assumptions}

We do not take into account the production point
distribution of neutrinos. This can be justified by two main reasons.
For the relevant values of $\Delta m^2$ our analysis shows that 
(see below) the resonance position lyes always outside the neutrino
production region ($r/R_{\rm SUN}<0.3$) because the solutions we find
implies $\Delta m^2 \lsim 10^{-7}$~eV$^2$. 
In order to see this we plot in Fig.~1 the matter potential as 
a function of the radial distance from the solar center. 

If neutrinos are considered Majorana particles, 
$V_{\rm matter} \equiv a_{\nu_e}-a_{\nu_\mu}$, 
and if they are Dirac, 
$V_{\rm matter} \equiv a_{\nu_e}$. 
It is shown, in the right ordinate in the same plot, also the value of 
the quantity $\Delta m^2$ to which a resonance is found. 
For example, if $\Delta m^2 = 10^{-7}$~eV$^2$ (for $E$= 1 MeV), 
a resonance is localized in $r/R_{\rm SUN}\approx 0.52$. 
Therefore, it can be seen that the resonance position is also
larger than $r/R_{\rm SUN} = 0.3$, outside the production region.  
Also, the range of $\mu_\nu B$ we consider here is much smaller than the matter
potential, $a_{\nu_e}$ and $a_{\nu_{\mu}}$, in the production region.
Again this fact is shown in Fig.~1, where the quantity $\mu_\nu B =
10^{-11}\mu_B\times 100$~kG is presented. 
Therefore, from eq. (\ref{evolution}) we observe that 
neutrino spin-flavor precession is prevented before neutrino gets 
the resonance region and the final survival probability is not 
affected by the production position.

It is obvious from the evolution equations (\ref{evolution}) that the
RSFP mechanism crucially depends on the solar magnetic field profile
along the neutrino trajectory. In the present paper we choose several
different profiles which, we believe, cover in general all the
previously~\cite{alp1,krastev,limnunokawa,pulido,chauhan} 
analysed magnetic profiles which led to a solution
to the solar neutrino anomaly. In Fig.~2, these magnetic fields are
presented in their general aspects. The constant magnetic profile
$B_1(r)$ was adopted in references~\cite{limnunokawa}, while the
general aspects of the profiles $B_3(r)$ and $B_4(r)$ have
already appeared in refs. \cite{krastev,pulido,alp1}, 
and \cite{chauhan}, respectively. 
We note that close to the solar surface ($r > 0.95 R_\odot$) we have
switched off the magnetic field for all the profiles  we considered
in this work.

\subsection{Definition of $\chi^2$}

The relevant free parameters in the RSFP mechanism are $\Delta m^2$ and
$\mu_\nu B$.  
Using the solar neutrino data given in Table I, 
we look for  the region in the $\Delta m^2 -\mu_\nu \vev{B}$ parameter space, 
where $\vev{B}$ denotes the average field strength defined as in Fig. 2, 
which leads to a solution to the solar neutrino anomaly by means
of the minimum $\chi^2$ method. 
In this analysis we will use only the SuperKamiokande data \cite{superkam}
without including the Kamiokande data \cite{kamiokande} 
because of the larger statistics and the smaller systematic 
error in the SuperKamiokande experiment. 
We also note that we will use the combined 
value of the two \ga experiments, 72.3 $\pm$ 5.6 SNU.

Our $\chi^2$ is defined as follows,
\begin{equation}
\chi^2 = \sum_{i,j}
(R_i^{th}-R_i^{obs})
[\sigma^2_{ij}(\mbox{tot})]^{-1}
(R_j^{th}-R_j^{obs}),
\end{equation}
where $(i,j)$ run through 
three experiments, i.e., \ga, \cl and SuperKamiokande, 
and the total error matrix $\sigma^2_{ij}(\mbox{tot})$ 
and the expected event rates $R_i$  are computed as follows.  
We essentially follow ref. \cite{FL} for the derivation of 
the error matrix and to describe the correlations of 
errors we used in this work. 
The expected experimental rates in the absence of 
neutrino conversion is given by, 
\begin{equation}
R_{i} = \sum_j C_{ij}\phi_j  \ \ \ 
(i= \mbox{Ga},\ \mbox{Cl},\  \mbox{SK}), 
\end{equation}
where $C_{ij}$ is the cross section coefficients and 
$\phi_j$ is the solar neutrino flux. 
In Sec. IV where we consider the case with neutrino conversion 
we use the coefficients $C_{ij}$ determined by properly 
convoluting the conversion probability in the integration 
over the neutrino energy spectrum for each detector and 
neutrino flux. 
In this work we consider neutrino from $pp$, $pep$, $^7$Be,  
$^8$B, $^{13}$N and $^{15}$O sources and neglect other minor 
flux such as $^{17}$F and $hep$ neutrinos.  

The total error matrix $\sigma^2_{ij}$ 
is the sum of the theoretical $\sigma^2_{ij}(\mbox{th})$ 
and experimental one $\sigma^2_{ij}(\mbox{exp})$, 
\begin{equation}
\label{total}
\sigma^2_{ij}(\mbox{tot}) = \sigma^2_{ij}(\mbox{th}) 
+ \sigma^2_{ij}(\mbox{exp}). 
\end{equation}
The theoretical error matrix can be further divided into 
the one coming from the uncertainties in the 
cross sections, $\sigma^2_{ij}(\mbox{cross})$  
and the one coming from 
uncertainties in the solar neutrino flux, 
$\sigma^2_{ij}(\mbox{flux})$,
\begin{equation}
\sigma^2_{ij}(\mbox{th}) = \sigma^2_{ij}(\mbox{cross}) 
+ \sigma^2_{ij}(\mbox{flux}). 
\end{equation}

The cross section error matrix $\sigma^2_{ij}(\mbox{cross})$ 
can be calculated by, 
\begin{eqnarray}
\sigma^2_{ij}(\mbox{cross}) &=& 
\delta_{ij}\sum_{k,l=1}^6 
\frac{\partial R_i}{\partial \mbox{ln}C_{kj} } 
\frac{\partial R_j}{\partial \mbox{ln}C_{lj} } 
\Delta \mbox{ln}C_{kj} \Delta \mbox{ln}C_{lj} \nonumber \\
& = & 
\delta_{ij}\sum_{k=1}^6 
(R_{ik}\Delta \mbox{ln}C_{ik})^2,
\end{eqnarray}
where $R_{ik}\equiv C_{ik} \phi_k$. 

On the other hand, the flux error matrix $\sigma^2_{ij}(\mbox{flux})$ 
can be calculated by, 
\begin{eqnarray}
\label{flux_error}
\sigma^2_{ij}(\mbox{flux}) &=& 
\sum_{k,l=1}^6 
\frac{\partial R_i}{\partial \mbox{ln}\phi_{k} } 
\frac{\partial R_j}{\partial \mbox{ln}\phi_{l} } 
\sum_{m=1}^{11} (\Delta \mbox{ln} \phi_{k})_m (\Delta \mbox{ln} \phi_{l})_m 
\nonumber \\
& = & 
\sum_{k,l=1}^6 
R_{ik}R_{jl}
\sum_{m=1}^{11} (\Delta \mbox{ln} \phi_{k})_m (\Delta \mbox{ln} \phi_{l})_m, 
\end{eqnarray}
where $(\Delta \mbox{ln} \phi_{k})_m$ is the fractional 
uncertainty of the $k$-th neutrino flux coming from the uncertainty 
in 
the $m$-th input parameter ($S_{11}$, $S_{33}$, $S_{34}$, $S_{17}$, 
$Z/X$, opacities, $S_{1,14}$, luminosity, age, diffusion or
$^7$Be + $e^{-}$ capture rate) which are obtained by the computer code 
exportrates.f, available at URL 
http://www.sns.ias.edu/$^\sim$jnb/. 
We note that in eq. (\ref{flux_error}) the product 
$(\Delta \mbox{ln} \phi_{k})_m (\Delta \mbox{ln} \phi_{l})_m$ takes
positive (negative) value when $k$-th and $l$-th fluxes are 
positively (negatively) correlated to each other 
with respect to the variation of the $m$-th input parameter. 

The experimental error matrix is given by, 
\begin{equation}
\sigma^2_{ij}(\mbox{exp}) = 
\delta_{ij} \sigma_i \sigma_j,
\end{equation}
where $\sigma_{i,j}$ ($i,j$ = Ga, Cl, SK) stands
for the combined error in each experiment. 

In Table II we show the correlation matrix $\rho_{ij}$ defined as,
\begin{equation}
\rho_{ij} \equiv 
\frac{ \sigma^2_{ij}(\mbox{tot}) }
{ \sqrt{\sigma^2_{ii}(\mbox{tot})\ \sigma^2_{jj}(\mbox{tot})} }.
\label{correlation}
\end{equation}
%
%
%
%
\begin{table}[h]
\caption[Tab]{The correlation matrix $\rho_{ij}$ obtained 
from eqs. (\ref{total})-(\ref{correlation}).}
\begin{center}
\begin{tabular}{cccc}
Experiment &  Correlation matrix &  &  \\ \hline
Ga         &   1.00  &   &    \\ 
Cl         &   0.497    & 1.00  &  \\
Super-Kam  &   0.486    & 0.952 & 1.00 \\
\end{tabular}
\end{center}
\label{tab:corr}
\vglue -3cm
\end{table}

\vglue 0.5cm 
\begin{figure}[ht]
\hglue 1.8cm
\epsfig{file=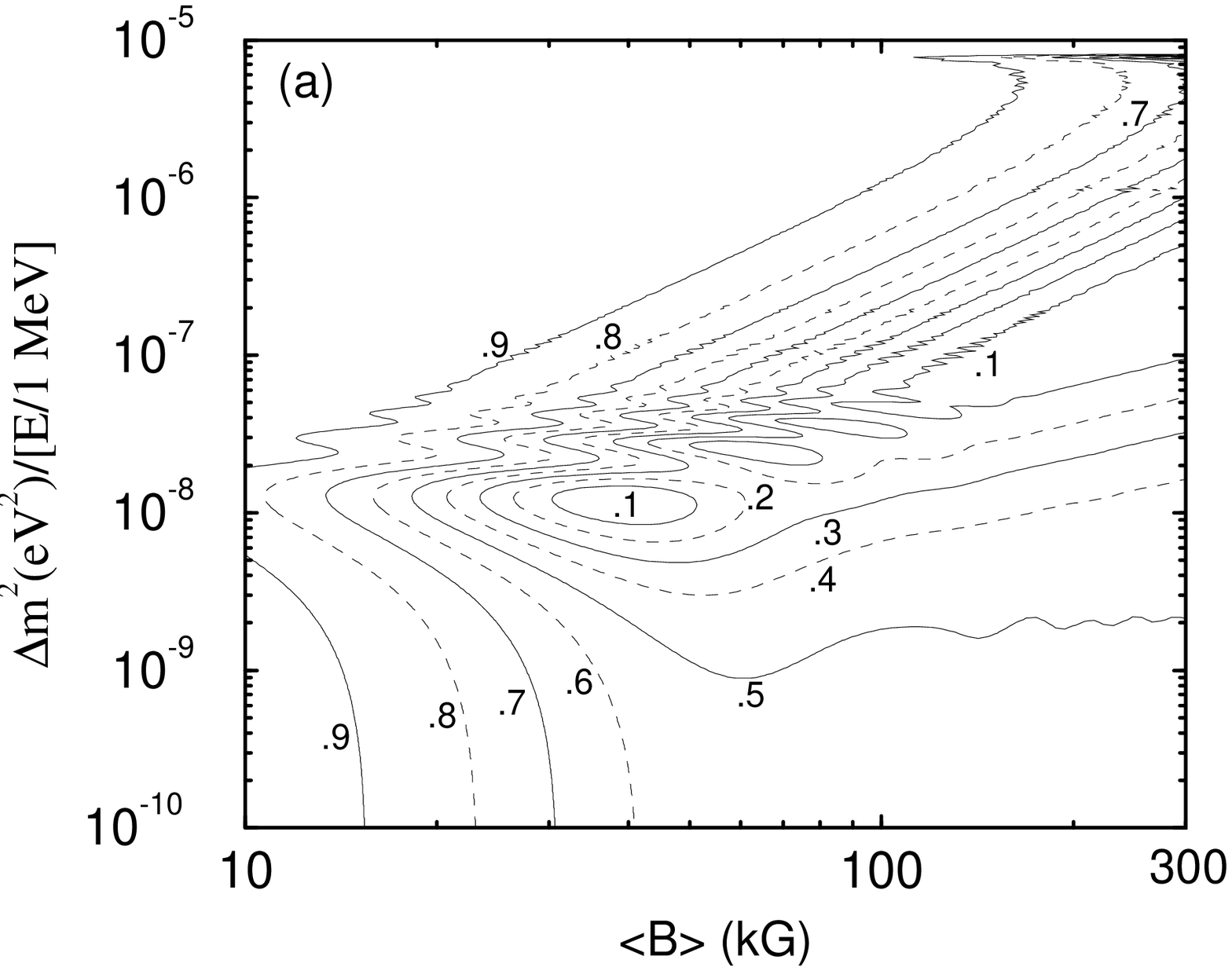,width=7.0cm}
\vglue -5.85cm \hglue 8.8cm \epsfig{file=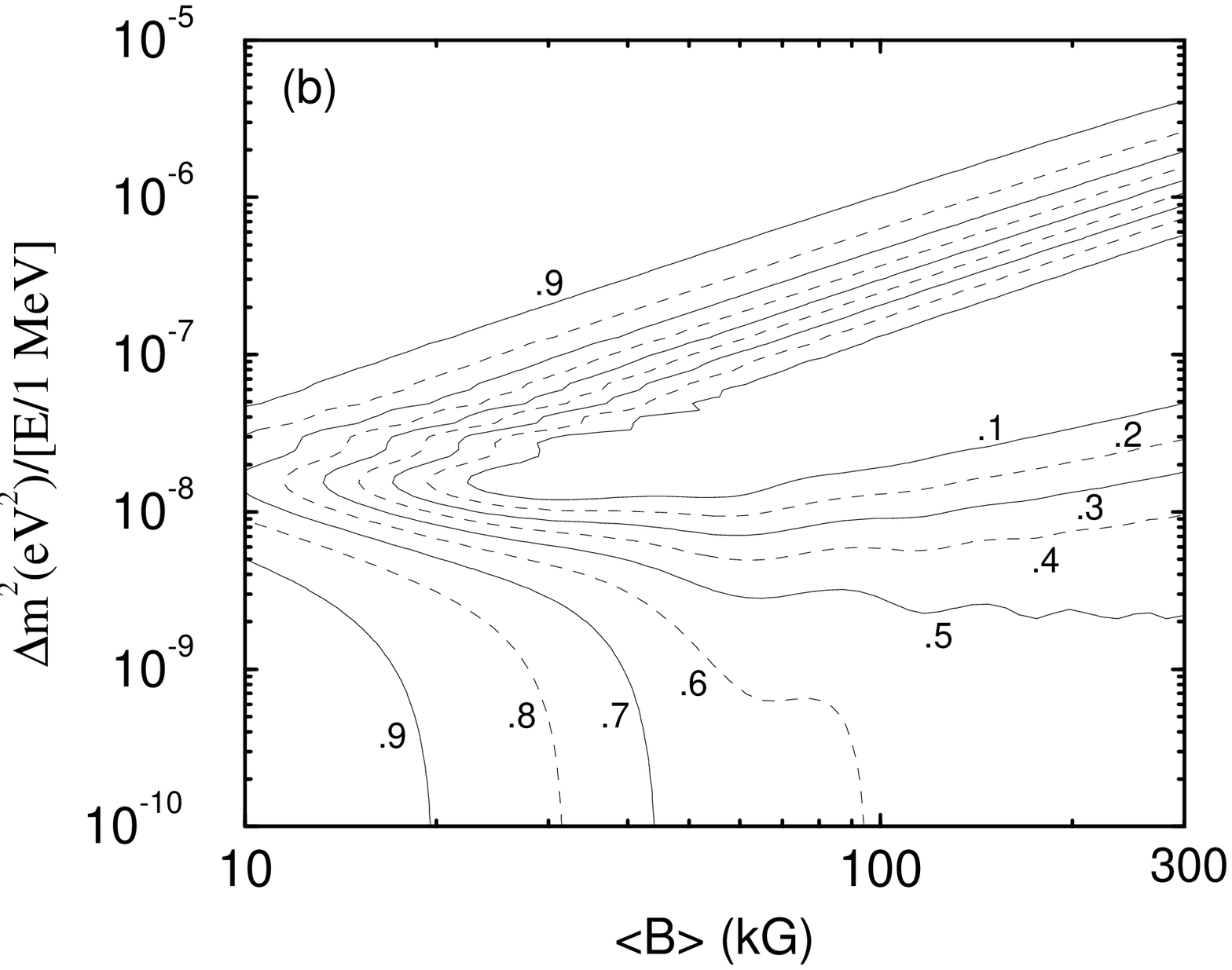,width=7.0cm}

\vglue -0.5cm 
\hglue 1.8cm
\epsfig{file=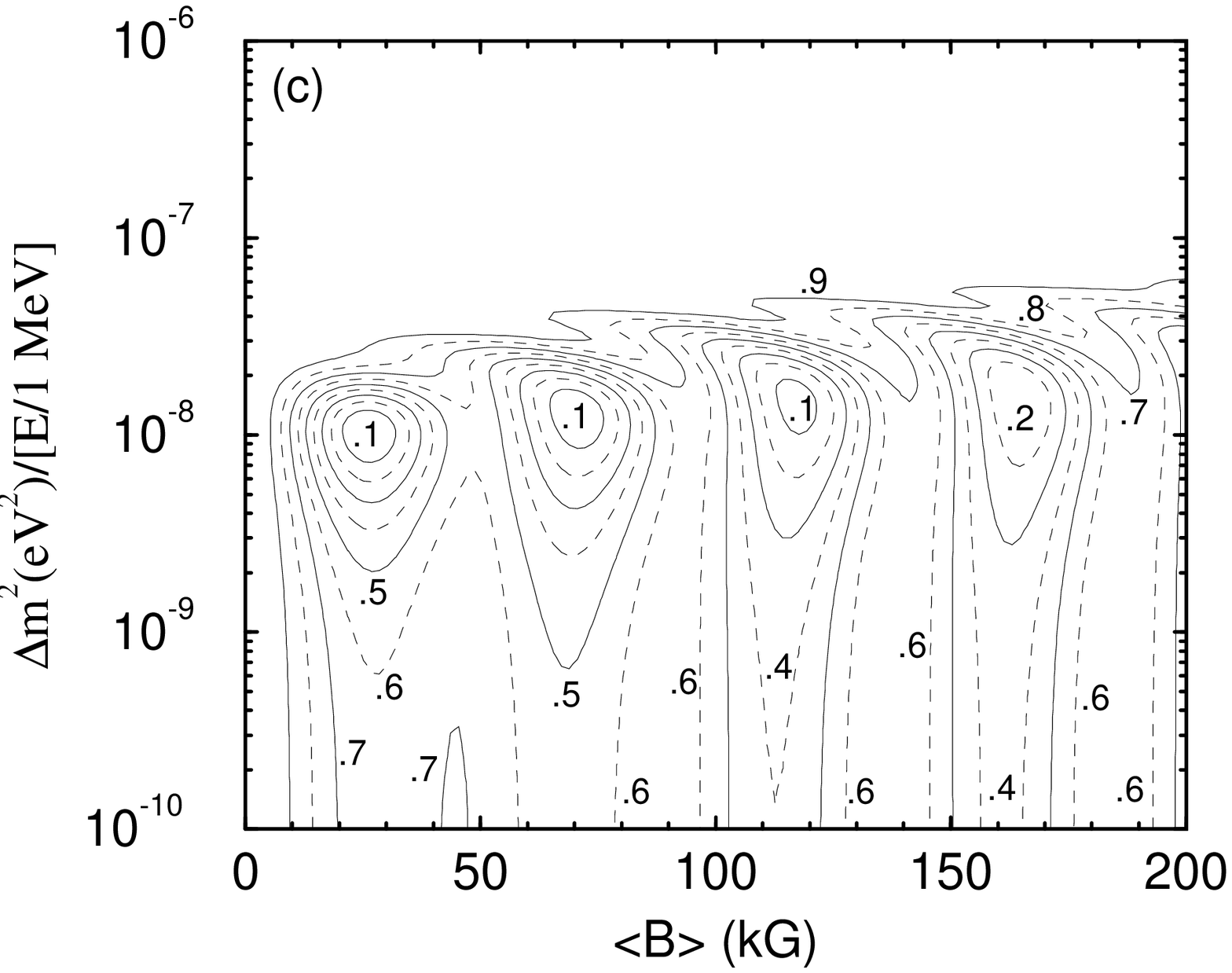,width=7.0cm}
\vglue -5.85cm \hglue 8.8cm \epsfig{file=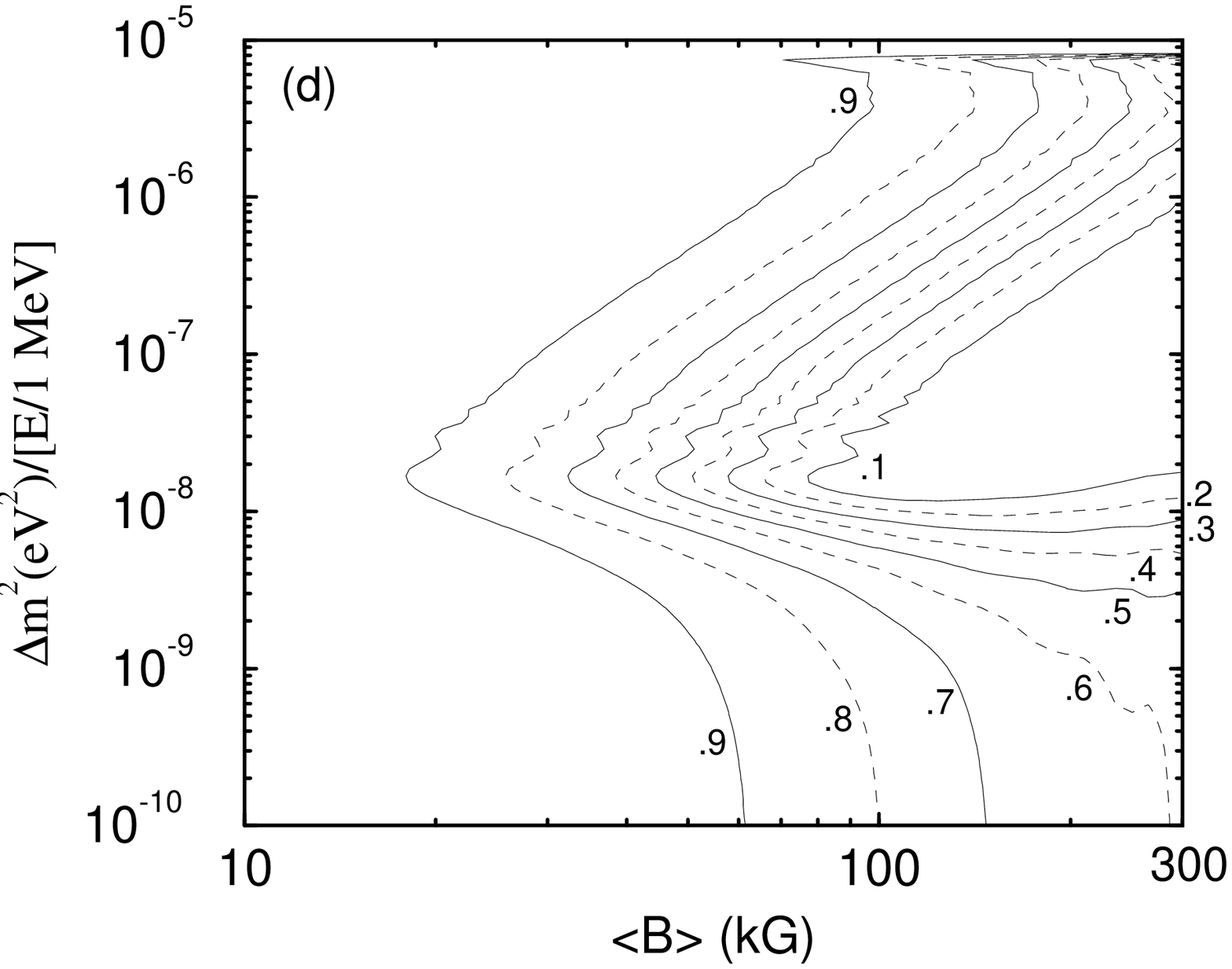,width=7.0cm}
\caption{
The contour plots of the survival probability 
in the $\vev{B}-\Delta m^2/E$ plane are shown 
in (a), (b), (c) and (d) for the magnetic field 
profiles, $B_1$, $B_2$, $B_3$ and $B_4$, respectively, 
sketched in Fig. 2. 
}
\end{figure}

\section{Results}

Now we compute the spin-flavor conversion probability, 
by numerically integrating the evolution eq. (1), assuming 
$\mu_\nu=10^{-11}\mu_B$ 
as a reference value, which is slightly below the
present experimental upper bound \cite{PDG98}. 
Hereafter we always assume this value of magnetic moment \cite{Raffelt} 
in this work but as it is clear from eq. (1) if  
$\mu_\nu$ is assumed to be smaller by certain amount, 
the same effect can still be obtained by simply increasing 
the value of magnetic field strength properly so that 
the product $\mu_\nu B$ would not be changed.

In Fig. 3 (a), (b), (c) and (d) we show the contour plots of 
the survival probability $P(\nu_e\to\nu_e)=
|\vev{\nu_e(t)|\nu_e(0)}|^2$ in the $\Delta m^2 - \vev{B}$ 
parameter space, for the magnetic profiles we sketched in Fig. 2, 
$B_1(r)$, $B_2(r)$, $B_3(r)$ and $B_4(r)$, respectively. 
We note that the field $B_3$ gives very different probability contours
from the other profiles, which will be also reflected in 
the final allowed region (see below).

Including now the experimental observations on
the solar neutrino signal above shown in Table~I, 
we can determine the region in the 
$\Delta m^2-\vev{B}$ parameter space which leads to a
RSFP solution to the solar neutrino problem  for a specified 
confidence level. 
We present the $\Delta m^2-\vev{B}$ parameter region 
which can account for all the solar neutrino data,  
at 90, 95 and 99 \% C.\ L. in Figs.~4 (a), (b), (c) and (d), 
for the magnetic profiles 
$B_1(r)$, $B_2(r)$, $B_3(r)$ and $B_4(r)$, respectively.

We observe from Figs.~4 (a) to (d) that a solution to the solar neutrino
problem can be found when $\vev{B} \gsim $ few times 10 kG 
and $\Delta m^2$ is the order of 
$10^{-8}$ to $10^{-7}$ eV$^2$ for any of the magnetic profiles 
used in this work.
Nevertheless, the quality of the fit, measured by the minimum $\chi^2$
criterion, varies a lot. The poorest fit is obtained when  the
continuously decaying magnetic field profile $B_4$ is used, with
$\chi^2_{min}=6.1$ for one (three data points - two free parameters) 
degrees of freedom.  Better fits are obtained when the $B_1$ (uniform) 
and $B_2$ (large triangle) fields are employed showing
$\chi^2_{min}=2.0$ and 1.8, respectively. And the best fit appears
when the triangular field in the solar convective  zone,  $B_3$, 
is employed, with a rather small value $\chi^2_{min}=0.13$.  
For this profile, we note that, as expected from Fig. 3 (c) 
we have several local best fit points also indicated in Fig. 4 (c) 
by the open circles whose corresponding $\chi^2_{min}$ are, from 
left to right,  2.3, 0.29 and 0.19.
%
%
\vglue 0.5cm 
\begin{figure}[ht]
\hglue 1.8cm
\epsfig{file=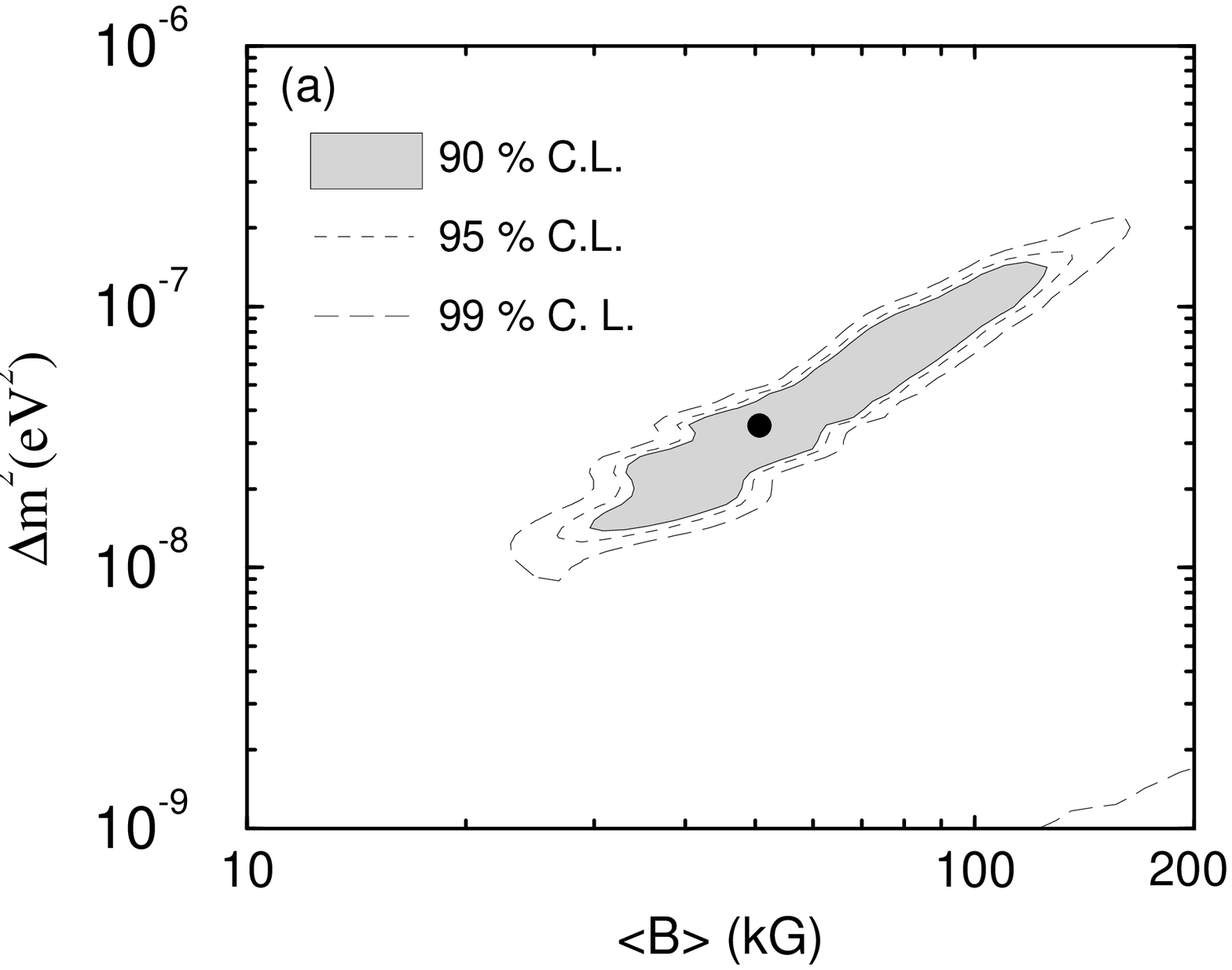,width=7.0cm}
\vglue -5.8cm \hglue 8.8cm \epsfig{file=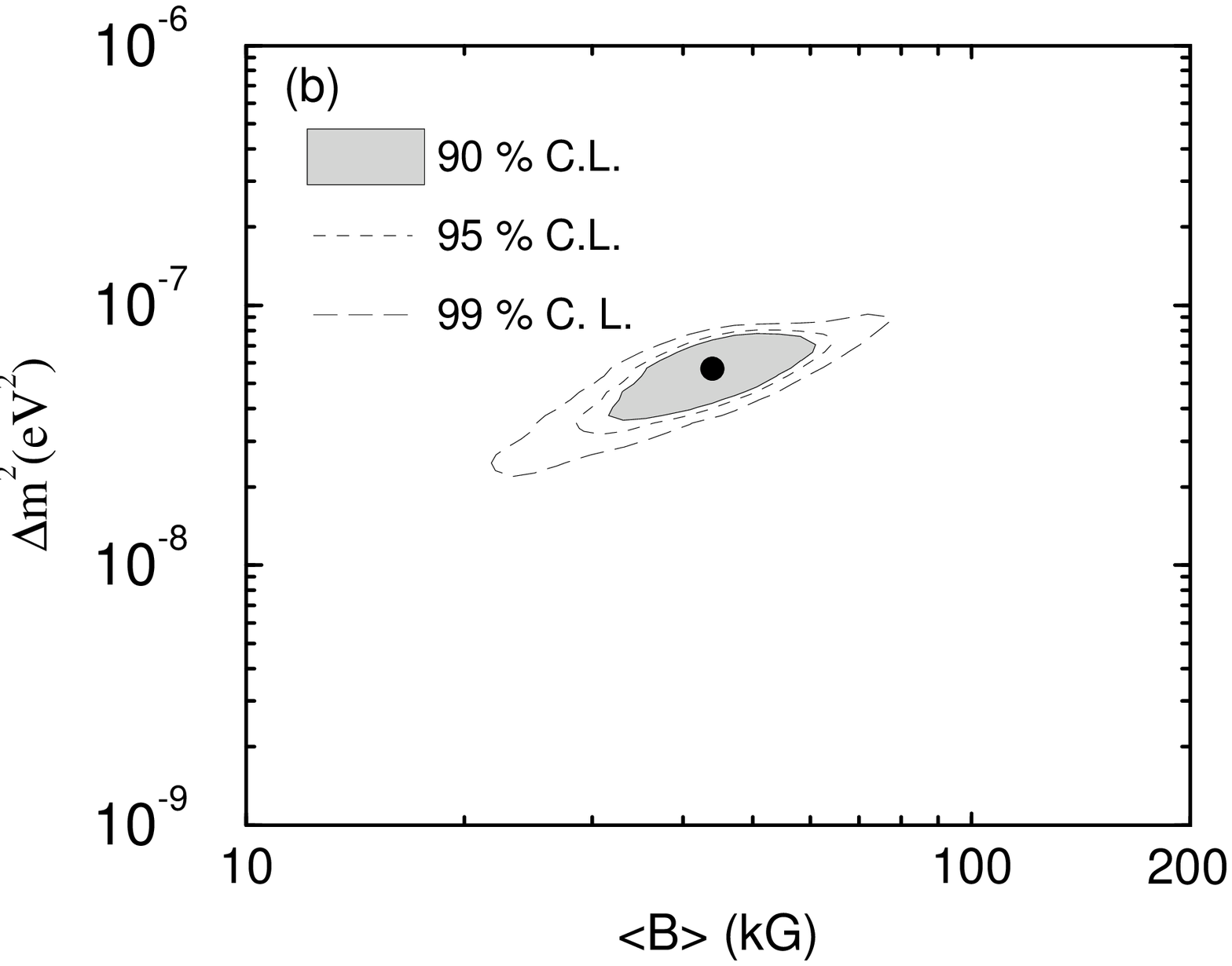,width=7.0cm}

\vglue -0.5cm 
\hglue 1.8cm
\epsfig{file=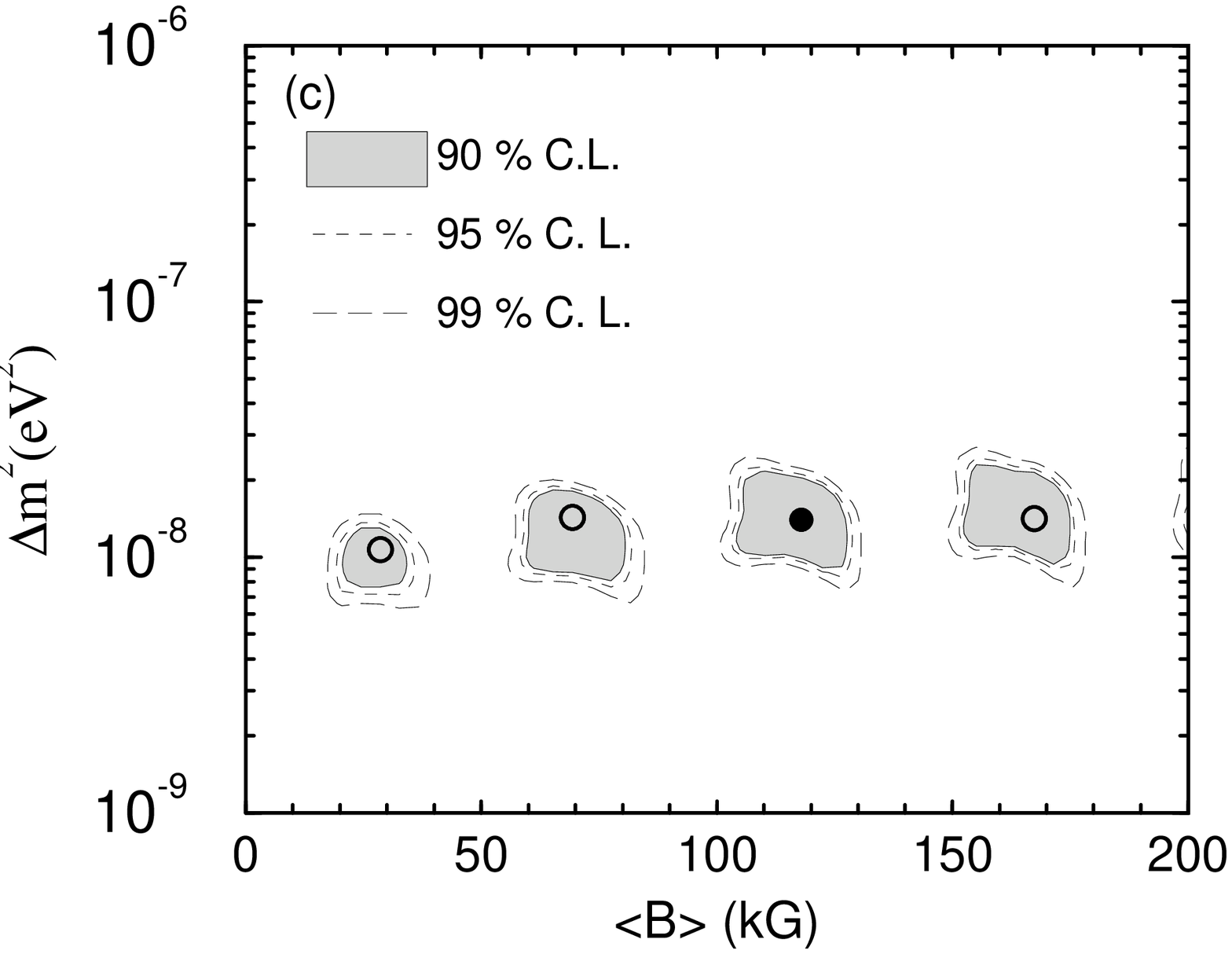,width=7.0cm}
\vglue -5.7cm \hglue 8.8cm \epsfig{file=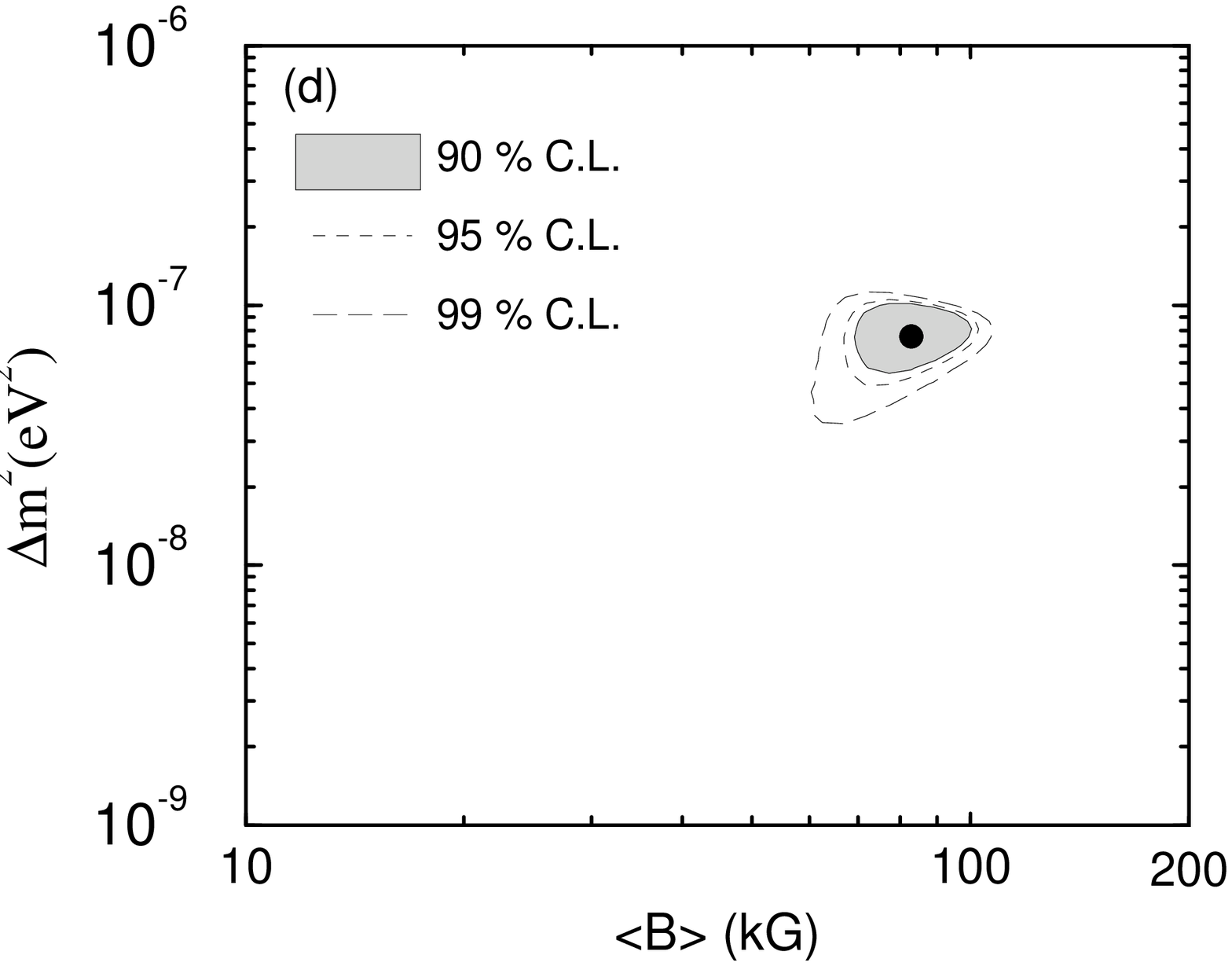,width=7.0cm}
\caption{
The Allowed RSFP solution to the solar neutrino problem. 
The parameter region allowed at 90, 95 and 99 \% C. L. 
are shown in (a), (b), (c) and (d) for the magnetic field 
profiles, $B_1$, $B_2$, $B_3$ and $B_4$, respectively, 
sketched in Fig. 1. 
We indicate best fit points by filled circles. 
In (c) we also indicate, by the open circles, 
the local best fit points inside each island 
delimited by 90 \% C.L. curves.  
}
\vglue 0.5cm 
\end{figure}

The reason why we are getting very good fit for $B_3$ is 
that this profile can provide the required suppression patterns 
of various neutrino flux implied by latest the data \cite{minakata98} 
as discussed in ref. \cite{pulido}. 
First we note that low energy $pp$ neutrinos are not so 
suppressed because the resonance positions are located in 
the inner region where the magnetic field is zero or small 
(see Figs. 1 and 2). 
However, intermediate energy \be7 neutrinos can be strongly 
suppressed due to the rapid increase of the field at the bottom 
of the convective zone since 
their resonance position is in a slightly outer region than 
the $pp$ one. 
On the other hand, high energy \b8 neutrinos are moderately 
suppressed because their resonance positions
are closer to the solar surface than the \be7 ones, where 
the field is decreasing.  
The best fitted values of $\vev{B}$ and $\Delta m^2$ as well 
as $\chi^2_{min}$ obtained from these different profiles
are summarized in Table III. 

\begin{table}[h]
\caption[Tab]{The best fitted parameters and $\chi^2_{min}$ for the 
Majorana case. Dirac case is presented in the parentheses.}
\begin{center}
\begin{tabular}{cccc}
Profile & $\vev{B}$ (kG) 
& $\Delta m^2$ (10$^{-8}$ eV$^2$) & $\chi^2_{min}$  \\ \hline
1       &   50.6\ (40.9)   &3.5\ (2.3)   &  2.0\ (6.2)  \\ 
2       &   47.1\ (40.8)   & 6.1\ (4.6)  &  1.8\ (5.7)  \\ 
3       &   118\ (69.4)   & 1.5\ (1.2)  & 0.13\ (1.3)  \\ 
4       &   82.9\ (81.6)   & 8.1\ (6.6) & 6.1\ (11.4)  \\ 
\end{tabular}
\end{center}
\label{tab:chi2}
\end{table}

We have repeated the same analysis also for 
the Dirac neutrino case. 
We, however, do not show the plots for the allowed region since they 
are rather similar to what have been presented above, 
if the same magnetic field profile is assumed. 
Instead, for the case of Dirac neutrinos, we only present 
the best fitted parameters and $\chi^2_{min}$ in 
the parentheses in Table III. 

We see from this table that, the Dirac case always leads 
to a worse fit if the same magnetic field profile is assumed. 
To understand this we should note that 
for the Dirac case, $\nu_e$'s are 
converted into the right handed muon (or tau) neutrino, 
which do not contribute to any of the solar experiments 
including the water Chrenkov experiment \cite{note}. 
In contrast in the Majorana case, converted right handed 
neutrino $\bar{\nu}_\mu$'s do 
contribute to the signal observed in the SuperKamiokande detector. 
This makes it difficult to conciliate the difference between 
the SuperKamiokande and \cl data. 

Let us now comment about the possibility of having 
such strong magnetic field in the Sun. 
While there is no generally accepted theory of solar 
magnetic field, it is possible to bound the field 
strength from very general arguments. 
It can be shown \cite{JBahcall} that the magnetic field less than 
10$^6$ kG in the solar core or less than 10$^4$ kG in the solar 
convective zone, will hardly affect the thermal structure and 
nuclear reaction processes well described by the standard solar model. 
These values come from the requirement that the magnetic pressure 
should be much smaller than the gas pressure, 
and can be regarded as the most generous upper limits of 
the magnetic field inside the Sun.  
More stringent bounds on the magnetic field in the convective zone 
are found in refs. \cite{SR,shi} where the discussion is 
based on the non-linear effects which eventually prevents the growth of 
magnetic fields created by the dynamo process. 
Naive limit can be obtained by estimating the required field tension
necessary to prevent a fluid element from sinking into a magnetically 
stratified region, so that the magnetic flux would not be further 
amplified. 
By equating the magnetic tension to the 
energy excess of a sinking element at the bottom of the
convective zone, Schmitt and Rosner \cite{SR} 
obtained $\sim 10$ kG as an upper bound for the magnetic field, 
which is of the order of the magnitude we need to have 
a good fit to the solar data by RSFP mechanism for the reference value 
of magnetic moment, $\mu_\nu=10^{-11}$ $\mu_B$. 

Finally, we briefly discuss how the recoil electron 
energy spectra in the SuperKamiokande detector 
will be affected by the RSFP mechanism \cite{pulido98}. 
In Fig. 5 (a) we plot the electron neutrino survival 
probabilities as a function of neutrino energy using 
the best fit parameters. 
In Fig. 5 (b) we plot the recoil electron energy 
spectra divided by the standard prediction 
expected to be observed in the SuperKamiokande detector, 
using also the best fit parameters as in Fig. 5 (a). 
In Fig. 5 (b) we also plot the latest data from 
SuperKamiokande \cite{superkam}. 
As we can see from the plot the observed data indicate some distortion 
mainly due to the last three data points in the higher energy bins. 
We, however, note from this plot that it seems difficult to 
exclude, at this moment, any of our predicted spectra expected 
from different field profiles, because of the experimental errors. 
We have to wait for more statistics and more careful analysis from 
the experimental group before drawing any definite conclusion. 
%

\begin{figure}[ht]
\hglue 1.4cm
\epsfig{file=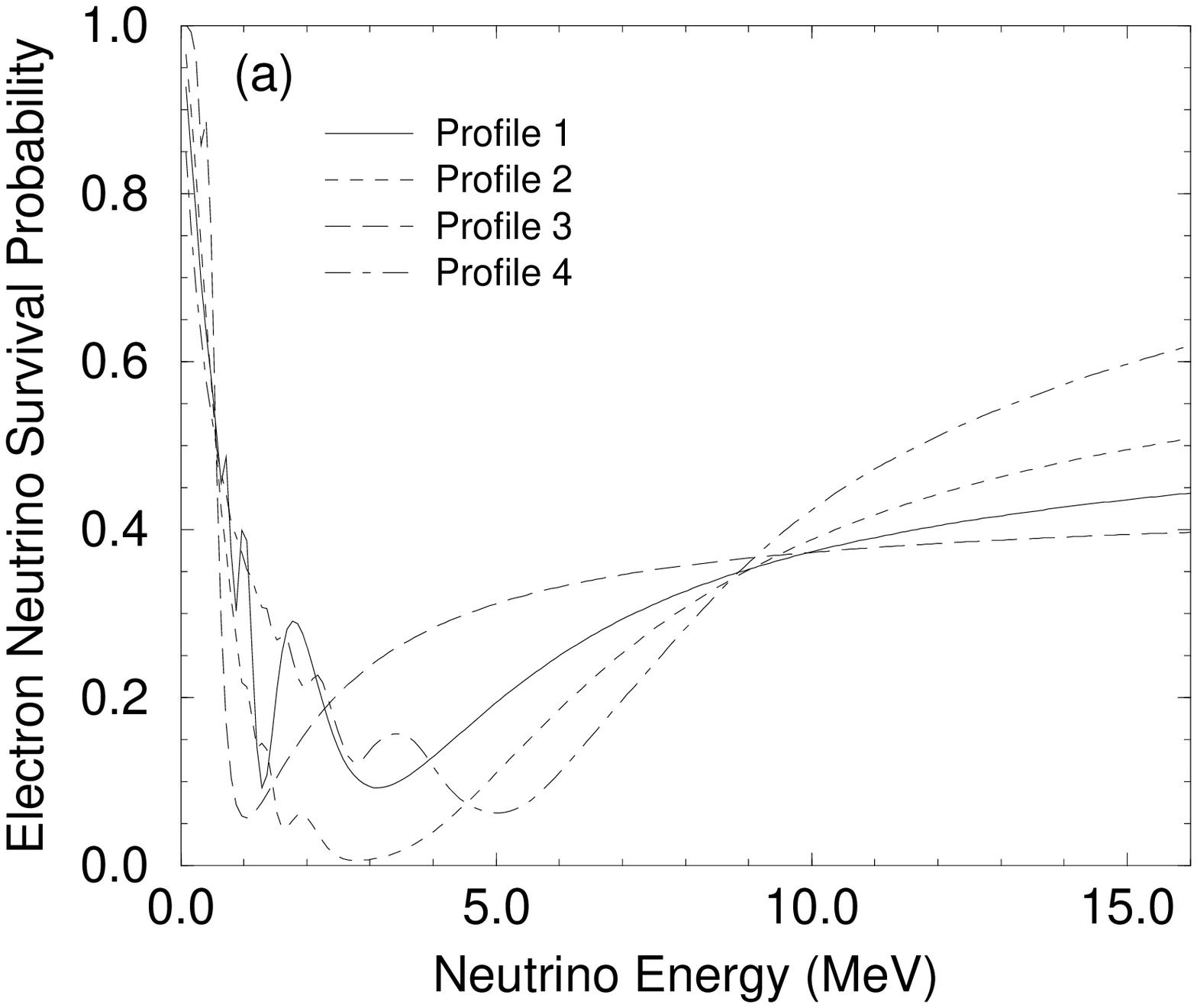,width=7.5cm}
\vglue -6.2cm \hglue 8.8cm \epsfig{file=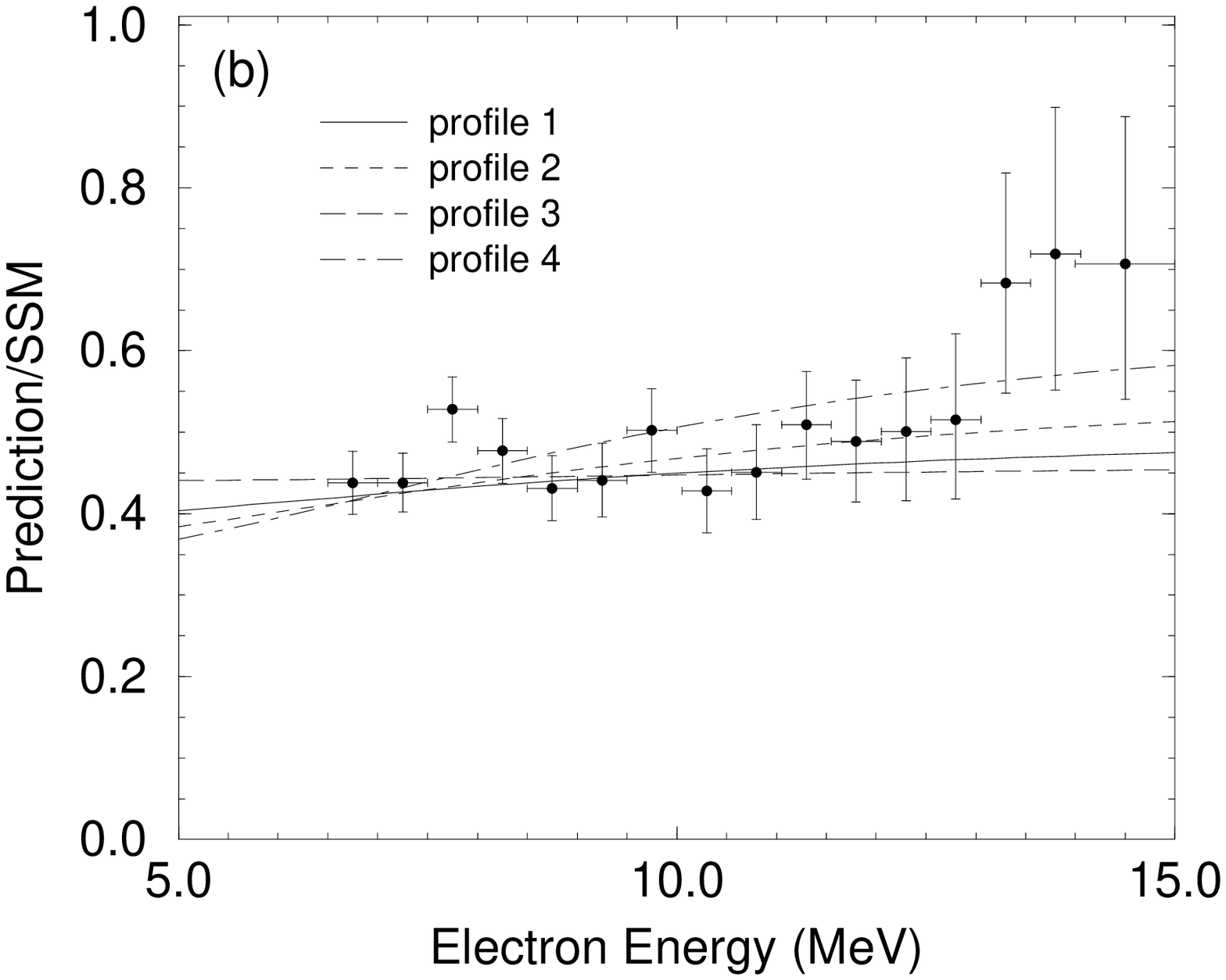,width=7.5cm}
\caption{We plot in (a) electron neutrino survival probability as a 
function of energy with the best fitted 
parameters for various field profiles. 
In (b) we plot recoil electron energy spectra expected from RSFP scenario 
using our best fit parameters, divided by the SSM prediction.  
The SuperKamiokande data are also shown by the filled circles with error bars. 
The last data point includes the contribution from the electrons with energy 
larger than 14 MeV. 
\label{fig:recoil}}
\end{figure}

\section{Conclusions}

We have reanalysed the RSFP mechanism as a solution to the solar neutrino 
problem in the light of the latest experimental data as well as 
the theoretical predictions. 

We found that the quality of the RSFP solution to the solar neutrino 
anomaly crucially depends on the solar magnetic field configuration 
along the neutrino trajectory inside the Sun. 
We found that the best fit to the observed solar neutrino data, 
which seems to be even better than the usual MSW solution as far 
as the total rates are concerned, is obtained if intensive magnetic 
fiels in the convective zone is assumed, in agreement with the conclusion 
found in ref. \cite{pulido}, whereas the linearly decaying magnetic 
field gives the worst fit. 
We, however, note that the required magnitude of the free parameters 
involved in the process, i.e., the magnetic field strength multiplied 
by the neutrino magnetic moment $\mu_\nu \vev{B}$ and the squared mass 
difference $\Delta m^2$, points to the same order, 
$\mu_\nu \vev{B} \approx \mbox{few\ times\ } 10^{-11}\mu_B \cdot 10$~kG 
and $\Delta m^2 \approx$ few times  $10^{-8}$~eV$^2$, for 
any of the field profiles assumed in this work. 

Our ignorance about the profile as well as the magnitude of the solar magnetic 
field makes this approach to the solar neutrino observation less predictive 
than its alternative approaches~\cite{vacuum,msw,fcnc1}. 
Nevertheless the presence of this mechanism opens
some interesting possibilities. 

One possibility is to look for any time variation of the solar 
neutrino signal \cite{mhd} which can not be expected in other
alternative solutions found in refs. \cite{vacuum,msw,fcnc1}. 
Any time variation of the solar neutrino signal which can be attributed 
to some time variation of the solar magnetic field can be a 
good signature of this mechanism. Although SuperKamiokande
has not yet confirmed any significant time variation 
up to experimental uncertainty this possibility remains. 

Another possibility is to look for the solar $\bar{\nu}_e$ flux, 
which can not be produced in the usual MSW or vacuum oscillation 
case but can be produced in RSFP mechanism if the flavor mixing is 
included. 
$\bar{\nu}_\mu$ produced by RSFP mechanism can be
converted into $\bar{\nu}_e$ by the usual vacuum oscillation. 
Ref. \cite{fiorentinie} suggests to observe (or to put upper bound of)
$\bar{\nu}_e$ flux 
in the SuperKamiokande whereas ref. \cite{pastor} 
suggests to use low energy solar neutrino experiment such 
as Borexino or Hellaz.  

We finally stress that RSFP mechanism can still provide a good solution 
to the solar neutrino problem, comparable in quality to MSW or 
Just So solution, and is not excluded by the present 
solar neutrino data. 

\vskip 0.5cm
\centerline{\bf Acknowledgements}
The authors would like to thank Eugeni Akhmedov and Jo\~ao Pulido 
for useful discussions and valuable suggestions, 
John Bahcall for helpful correspondence and encouragement, 
Andrei Gruzinov for the helpful comment regarding to the size 
of the solar magnetic field, Eligio Lisi for useful comments 
regarding to the $\chi^2$ analysis. 
The authors would also like to thank Conselho Nacional de Desenvolvimento
Cient\'\i fico e Tecnol\'ogico  (CNPq), PRONEX 
and Funda\c{c}\~ao de Amparo \`a Pesquisa do Estado de S\~ao Paulo (FAPESP) 
for several financial supports.  
H. N. has been supported by a postdoctral fellowship from FAPESP.


\begin{thebibliography}{99}


\bibitem{homestake}
K.\ Lande (Homestake Collaboration) in 
{\it Neutrino '98}, Proceedings of the XVIII International Conference on
Neutrino Physics and Astrophysics, Takayama, Japan, 4--9 June 1998,
edited by Y. Suzuki and Y. Totsuka, 
to be published in Nucl. Phys. B (Proc. Suppl.), 
scanned transparencies are available at 
URL http://www-sk.icrr.u-tokyo.ac.jp/nu98/scan/index.html.
 
\bibitem{kamiokande}
Y. Fukuda {\sl et al.} (Kamiokande Collaboration), 
Phys. Rev. Lett. {\bf 77}, 1683 (1996). 

\bibitem{sage}
V. Gavrin (SAGE Collaboration) in {\it Neutrino '98} \cite{homestake}. 

\bibitem{gallex}
T. Kirsten (GALLEX Collaboration) in {\em Neutrino '98} \cite{homestake}. 


\bibitem{superkam}
Y. Suzuki (SuperKamiokande Collaboration) 
in {\it Neutrino '98} \cite{homestake}. 

\bibitem{JBahcall} J. Bahcall and R. Ulrich,  Rev. of Mod. Phys. {\bf
60}, 297 (1988); J. N. Bahcall, {\it Neutrino Astrophysics}, Cambridge
University Press, New York, (1989); the neutrino fluxes as well 
as cross sections are available at URL http://www.sns.ias.edu/~jnb/. 

\bibitem{BP95}
J. N. Bahcall and M.H. Pinsonneault, 
Rev. of Mod. Phys. {\bf 67}, 781 (1995).

\bibitem{BP98}
J. N. Bahcall, S. Basu and M.H. Pinsonneault, 
Phys. Lett. B {\bf 433}, 1 (1998). 

\bibitem{reviewSSM}
For a recent review, see J.\ N.\ Bahcall, astro-ph/9808162.

\bibitem{INT}
E. G. Adelberger {\sl et al.}, astro-ph/9805121, 
Rev. Mod. Phys.,  {\bf 70}, 1265 (1998). 

\bibitem{BKS}
J. N. Bahcall, P. I. Krastev and A. Yu. Smirnov,
Phys. Rev. {\bf D58}, 096016 (1998). 

\bibitem{minakata98}
H. Minakata and H. Nunokawa, Phys. Rev. {\bf D59}, 073004 (1999), 
and references therein for the previous works.  

\bibitem{30years} For a detailed list of references 
on these solutions see also, 
Solar Neutrinos: The First Thirty Years, 
ed. by R. Davis Jr. {\sl et al.},
Frontiers in Physics, Vol. 92, 
Addison-Wesley, 1994. 


\bibitem{vacuum}
V. N. Gribov and B. M. Pontecorvo, 
Phys. Lett. B {\bf 28}, 493 (1969). 

\bibitem{msw}
S. P. Mikheyev and A. Yu. Smirnov,  Sov. J. Nucl. Phys.{\bf 42},
913 (1985); Nuovo Cimento {\bf C9}, 17 (1986);
L. Wolfenstein,   Phys. Rev. {\bf D17}, 2369 (1978).

\bibitem{LimMarciano}
C. S. Lim and W. J. Marciano,  Phys. Rev. {\bf 37}, 1368 (1988).

\bibitem{Akhmedov}
E. Kh. Akhmedov,  Sov. J. Nucl. Phys. {\bf 48}, 382 (1988);
Phys. Lett. {\bf B213}, 64 (1988).

\bibitem{fcnc1}
M. M. Guzzo, A. Masiero and S. T. Petcov, Phys. Lett. {\bf B260}, 154 (1991);
V. Barger, R. J. N. Phillips and K. Whisnant, 
Phys. Rev. {\bf D44},1692 (1991);
J. Bahcall and P. Krastev, hep-ph/9703267.

\bibitem{fcnc2}
E. Roulet, Phys. Rev. {\bf D44}, 935 (1991). 

\bibitem{update}
See for e.g, 
N. Hata and P. Langacker, Phys. Rev. {\bf D56}, 6107 (1997);
G. L. Fogli, E. Lisi and D. Montanino, Astropart. Phys. {\bf 9}, 119 (1998);
J. N. Bahcall, P. I. Krastev and A. Yu. Smirnov, in ref. \cite{BKS};
P.\ I.\ Krastev and J.\ N.\ Bahcall in \cite{fcnc1}. 


\bibitem{review}
For reviews, see for e.g., J. Pulido, Phys. Rep {\bf 211}, 167 (1992);
E. Kh. Akhmedov,  talk presented at 4th International Solar 
Neutrino Conference, Heidelberg, Germany, April 1997, 
hep-ph/9705451 and references therein. 

\bibitem{alp1}
E. Kh. Akhmedov,  A. Lanza,  and S. T. Petcov, 
Phys. Lett. {\bf B303},  85 (1993).

\bibitem{krastev}
P. I. Krastev,  Phys. Lett. {\bf B303}, 75 (1993).

\bibitem{limnunokawa}
C. S. Lim and H. Nunokawa, 
Astropart. Phys. {\bf 4}, 63 (1995). 


\bibitem{pulido}
J. Pulido, Phys. Rev. {\bf D57}, 7108 (1998).

\bibitem{chauhan}
B. C. Chauhan, U. C. Pandey and S. Dev,
Mod. Phys. Lett. {\bf A13}, 1163 (1998). 

\bibitem{cisneros}
A. Cisneros, Astrophys. Space Sci. {\bf 10}, 87 (1971). 

\bibitem{vvo}
L. B. Okun, M. B. Voloshin,  and M. I. Vysotsky,  Sov. Phys. JETP
{\bf 64}, 446 (1986).

\bibitem{schechtervalle}
J. Schechter and J. W. F. Valle, 
Phys. Rev. {\bf 24}, 1883 (1981); 
{\sl ibid}{\bf 25}, 283 (1982).


\bibitem{rsfpmsw}
H. Minakata and H. Nunokawa, Phys. Rev. Lett. {\bf 63}, 121 (1989);
%
A. B. Balantekin, P. J. Hatchell and F. Loreti,  
Phys. Rev. {\bf D41}, 3583 (1990). 
%
H. Nunokawa and H. Minakata,  Phys. Lett. {\bf B314}, 371 (1993);
%
E. Kh. Akhmedov,  A. Lanza,  and S. T. Petcov,
Phys. Lett. {\bf B348}, 124 (1995). 

\bibitem{BKL}
J. N. Bahcall, P. I. Krastev and E. Lisi, Phys. Rev. {\bf C55}, 494 (1997). 

 

\bibitem{FL}
G. L. Fogli and E. Lisi, Astro. Part. Phys. {\bf 3}, 185 (1995).


\bibitem{PDG98} C.\ Caso {\em et al.}, The European Physical Journal 
{\bf C3}, 1 (1998). 

\bibitem{Raffelt}
We note, however, that the more stringent bound, 
$\mu_\nu < 3\times 10^{-12}\mu_B$, is obtained from the 
argument of the cooling of the red giants, 
in G. Raffelt,  Phys. Rev. Lett. {\bf 64}, 2856 (1990).

\bibitem{SR}
J. Schmitt and R. Rosner, Astrophys. J. {\bf 265}, 901 (1983). 

\bibitem{shi}
X. Shi {\em et al.}, Comm. Nucl. Part. Phys. 
{\bf 21}, 151 (1993). 


\bibitem{note}
Note that, strictly speaking, such right handed 
muon (or tau) neutrinos can contribute
to water Cherenkov detector though electromagnetic 
interaction provided that $\mu_\nu$ is large enough. 
However, such contribution is small enough 
to be neglected for the magnetic moment we 
are assuming in this work. 

\bibitem{pulido98}
J. Pulido, hep-ph/9808319. 

\bibitem{mhd} M. M. Guzzo, N. Reggiani and J.H.Colonia, 
Phys. Rev. {\bf D56}, 588 (1997); 
M. M. Guzzo, N. Reggiani, P. H. Sakanaka, Phys. Lett. {\bf B357}, 602 (1995).  

\bibitem{fiorentinie} 
G. Fiorentini, M. Moreti and F.L.Villante, Phys. Lett. {\bf B413}, 378 (1997). 

\bibitem{pastor} 
S. Pastor, V.B.Semikoz and J.W.F.Valle, Phys. Lett. {\bf B423}, 118 (1998). 


\end{thebibliography}
\end{document}